\documentclass[12pt]{article}
\newcommand{\bydef}{\stackrel{\mathrm{def}}{=}}

\usepackage{amssymb,graphicx}

\title{Protected qubit based on a superconducting current mirror}

\author{Alexei Kitaev\\
\normalsize\it
Microsoft Project Q, UCSB, Santa Barbara, California 93106, USA\\
\normalsize\it
California Institute of Technology, Pasadena, California 91125, USA}

\date{September 19, 2006}

\begin{document}

\maketitle

\begin{abstract}
We propose a qubit implementation based on exciton condensation in
capacitively coupled Josephson junction chains. The qubit is protected in the
sense that all unwanted terms in its effective Hamiltonian are exponentially
suppressed as the chain length increases. We also describe an implementation
of a universal set of quantum gates. Most gates also offer exponential error
suppression. The only gate that is not intrinsically fault-tolerant needs to
be realized with ${\sim}\:50\%$ precision, provided the other gates are exact.
\end{abstract}


Physical implementation of a quantum computer presents a great challenge
because quantum systems are susceptible to decoherence and because
interactions between them cannot be controlled precisely. There has been
impressive demonstrations of qubits using various kinds of systems, including
Josephson
junctions~\cite{NPT,quantronium,Yu_at_al,large_qubit,flux_qubit,Jgate}, yet
building a full-scale computer remains a remote goal.  In principle,
scalability can be achieved by correcting errors at the logical
level~\cite{Shor}, but only if the physical error rate is sufficiently
small. As an alternative solution, it was suggested that topologically ordered
quantum systems are physical analogues of quantum error-correcting codes, and
fault-tolerant quantum computation can be performed by braiding
anyons~\cite{Kitaev}.

More recently, Doucot and Vidal~\cite{DoucotVidal}, and Ioffe and
Feigelman~\cite{IF02} found simpler examples of physical systems
with error-correcting properties. The key element of their proposals, which we
refer to as $0$-$\pi$ qubit, is a two-terminal circuit built of Josephson
junctions. Its energy has two equal minima when the superconducting phase
difference between the terminals, $\theta=\phi_1-\phi_2$ is equal to $0$ or
$\pi$. The quantum states associated with the minima, $|0\rangle$ and
$|1\rangle$ can form quantum superpositions. It is essential that that the
energy difference between the two minima is exponentially small in the system
size, even in the presence of various perturbations, hence the quantum
superposition will remain unchanged for a long time. Implementations of some
quantum gates were also proposed.

\begin{figure}
\centerline{\includegraphics[scale=0.8]{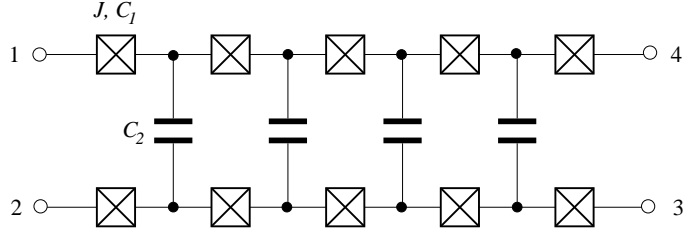}}
\caption{The superconducting current mirror.}
\label{fig_ladder}
\end{figure}

In this paper, we discuss a different design of a $0$-$\pi$ qubit. It is based
on the current mirror effect in capacitively coupled chains of Josephson
junctions, see Fig.~\ref{fig_ladder}. An analogue of this effect in
normal-metal junctions is due to correlated electron-hole
tunneling~\cite{AKN}, whereas in superconducting chains the tunneling objects
are Cooper pairs. Positive and negative Cooper pairs (with electric charge
$+2e$ in one chain and $-2e$ in the other chain) tend to tunnel
together. Under suitable conditions, the currents in the two chains are
opposite in direction and almost equal in magnitude. This was observed
experimentally~\cite{SD} in the resistive state, i.e., at sufficiently large
voltage bias. However, we will be concerned with a more delicate,
dissipationless form of this effect, which has not been observed yet but
predicted theoretically by Mahn-Soo Choi, M.\,Y.\,Choi, and Sung-Ik
Lee~\cite{CCL} for the case of strong interchain coupling. In this regime, the
Josephson junction ladder behaves as an almost perfect DC transformer with 1:1
current and voltage ratio.

Let each junction in the ladder have Josephson energy $J$ and capacitance
$C_1$, and let $C_2$ be the interchain capacitance. If $C_2\gg C_{1}$,
excitons consisting of $+2e$ in one chain and $-2e$ in the other chain have
lower energy than individual $\pm 2e$ quasiparticles or other excitations that
change the total charge on some rungs of the ladder. The energy scales for
excitons and unbalanced charges are given by $E_\mathrm{ex}\sim e^{2}/C_2$ and
$E_1\sim e^{2}/C_1$, respectively. Excitons form a Bose condensate if
$E_\mathrm{ex}\lesssim J_\mathrm{ex}$, where $J_\mathrm{ex}\sim J^2/E_1$ is a
characteristic hopping energy (we assume that $J\lesssim E_1$). In this
regime, the system becomes superconducting with respect to opposite currents
in the two chains while being insulating with respect to passing net electric
charge along the ladder. It is worth noting that the exciton condensate
persists in the presence of charge frustration~\cite{LCC}.

The current mirror device may be characterized by an effective potential
energy $E$ that depends on the values of the superconducting phase at the four
terminals, $\phi_1,\phi_2,\phi_3,\phi_4$. The order parameter of the exciton
condensate may be represented as the superconducting phase difference between
the chains, which is equal to $\theta_l=\phi_1-\phi_2$ at the left end of the
ladder and $\theta_r=\phi_4-\phi_3$ at its right end. Thus we expect the
energy to depend primarily on $\theta_l-\theta_r=\phi_1-\phi_2+\phi_3-\phi_4$:
\begin{equation}\label{mirror}
E=F(\phi_1-\phi_2+\phi_3-\phi_4)+ f(\phi_1-\phi_4,\phi_2-\phi_3),
\end{equation}
where $f$ is an ``error term''. Since the current through the $j$-th terminal
is proportional to $\partial E/\partial\phi_{j}$, the error term characterizes
the net current through the ladder. Such current can only be carried by $\pm
2e$ quasiparticles tunneling through the insulator, but this process is
suppressed by factor $\exp(-N/N_0)$, where $N$ is the length (the number of
junctions in each chain) and $N_{0}\sim 1$. On the other hand, the $F$ term in
Eq.(\ref{mirror}) is of the order of $J_\mathrm{ex}/N$.

Now we will explain the design of the $0$-$\pi$ qubit, which is very
simple. Let us connect the four leads diagonally, i.e., 1 with~3 and 2
with~4. Thus $\phi_1=\phi_3$,\, $\phi_2=\phi_4$, and $E\approx
F(2(\phi_1-\phi_2))$ with exponential precision. Since the function
$F(\theta)$ has a minimum at $\theta=0$, the energy of the qubit has two
minima: at $\phi_1-\phi_2=0$ and at $\phi_1-\phi_2=\pi$ (note that all the
variables $\phi_j$ are defined modulo $2\pi$). The energy values at the minima
are exponentially close to each other: $\delta E\propto\exp(-N/N_{0})$. That
is the reason for protection against dephasing. To prevent bit flips, one
needs to make sure that $E\gg e^{2}/C$, where $C=NC_2$ is total interchain
capacitance. Note that the ratio $E/(e^{2}/C)\sim J_\mathrm{ex}/E_{ex}$ does
not depend on the length. It can be increased by increasing the interchain
coupling or by connecting several current mirrors in parallel.

\begin{figure}
\centerline{\begin{tabular}{c@{\qquad\qquad}c}
\includegraphics[scale=0.8]{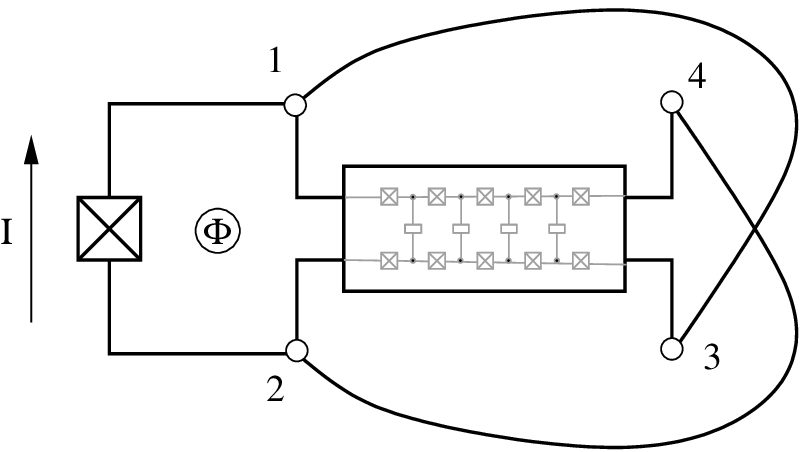} &
\includegraphics[scale=0.8]{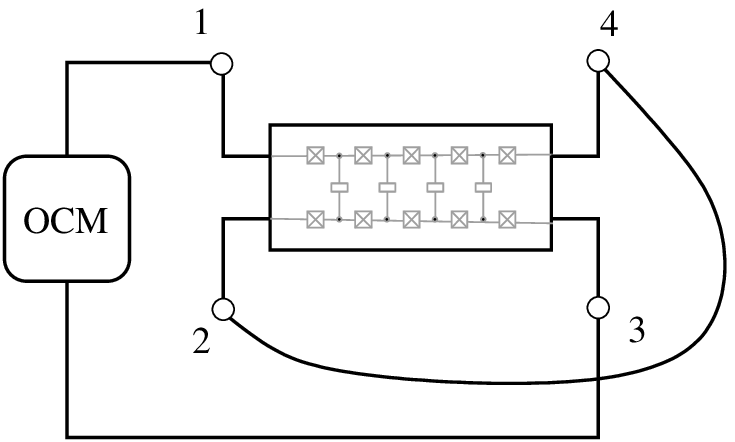} \\
a) & b)
\end{tabular}}
\caption{Measurement in the standard basis (a) and in the dual basis (b).
$\Phi$ designates magnetic flux, and OCM a circuit for the offset charge
measurement.}
\label{fig_meas}
\end{figure}

With this qubit design, it is possible to do measurements in the standard
basis (of states $|0\rangle$ and $|1\rangle$ corresponding to
$\phi_1-\phi_2=0$ and $\phi_1-\phi_2=\pi$, respectively) as well as in the
dual basis, $|\pm\rangle=\frac{1}{\sqrt{2}}(|0\rangle\pm |1\rangle)$. These
may also be called ``phase basis'' and ``charge basis''; the reason for the
second name will be clear later. The measurement in the phase basis can be
performed by simply connecting leads 1 and~2 to a measuring circuit. For
example, if the leads are connected via a Josephson junction, the current in
the loop depends on $\phi_1-\phi_2=0,\pi$ and the magnetic flux through the
loop,~Fig.~\ref{fig_meas}a.

The dual measurement is more complicated. The key idea is to break the wire
between 1 and~3 and attach some measuring circuit to those
leads,~Fig.~\ref{fig_meas}b. Now
\begin{equation}\label{charge_measurement}
\phi_2=\phi_4\equiv (\phi_1+\phi_3)/2 \pmod{\pi}.
\end{equation}
Furthermore, the potential energy is practically independent of the
superconducting phase difference $\theta=\phi_1-\phi_3$ across thus configured
device, hence direct superconducting current cannot flow. The device behaves
essentially as a capacitor with the effective Hamiltonian
\begin{equation}
H_{C}=\frac{(2e)^2}{2C}
\left(\frac{\partial}{i\,\partial\theta}-n_{g}\right)^{2},
\end{equation}
except that it has an internal degree of freedom, since for fixed values of
$\phi_1,\phi_3$ Eq.~(\ref{charge_measurement}) has two solutions. The states
$|+\rangle$ and $|-\rangle$ correspond to the symmetric and antisymmetric
superposition of these solutions, and the wave function $\psi(\theta)$
satisfies the boundary condition $\psi(2\pi)=\psi(0)$ or
$\psi(2\pi)=-\psi(0)$, respectively. The second boundary condition becomes
equivalent to the first one if we change $n_{g}$ by $1/2$. (The parameter
$n_{g}$ is a so-called offset charge measured in units of $2e$; it is defined
modulo $1$.) Thus, the measurement in the $|\pm\rangle$ basis amounts to
distinguishing between $n_{g}$ and $n_{g}+1/2$. From the practical
perspective, $n_{g}$ need not be known in advance. Indeed, it is only
important to tell the two states apart while the labels ``$+$'' and ``$-$''
can be assigned arbitrarily. We do not describe a concrete procedure for the
offset charge measurement but only remark that this is a general problem not
pertaining to the particular device.

The application of qubits for universal or specialized quantum computation
requires some implementation of quantum gates. It is very desirable for the
gates to be fault-tolerant at the physical level, or at least to limit
possible errors to some particular types. We now sketch such an
implementation. It involves the choice of a nonstandard but computationally
universal set of gates that are particularly suitable for the use with
$0$-$\pi$ qubits. The computation is adaptive, i.e., it proceeds by
intermediate measurements whose outcome determines the choice of the next gate
to be applied. A tentative realization of the gates will be described briefly,
with the intension to demonstrate the feasibility of the whole scheme.

The gates we use are as follows:
\begin{enumerate}
\item Measurement in the standard basis.
\item Measurement in the dual basis.
\item One-qubit unitary gate
$R(\pi/4)=\exp(i(\pi/4)\sigma^z)$ and its inverse.
\item Two-qubit unitary gate
$R_{2}(\pi/4)=\exp(i(\pi/4)\sigma^z_1\sigma^z_2)$ and its inverse.
\item One-qubit unitary gate
$R(\pi/8)=\exp(i(\pi/8)\sigma^z)$ and its inverse.
\end{enumerate}
To show that this set is universal, we first observe that repeated
applications of noncommuting measurements allow one to prepare any of these
states: $|0\rangle$, $|1\rangle$, $|+\rangle$, $|-\rangle$. Notice that
$R(-\pi/4)$ is equal to the commonly used operator $\Lambda(i)$ up to an
overall phase, where $\Lambda(i)|a\rangle=i^{a}|a\rangle$ for $a=0,1$. One can
also implement the two-qubit controlled phase gate $\Lambda^2(-1)$ that acts
as follows: $\Lambda^2(-1)|a,b\rangle=(-1)^{ab}|a,b\rangle$. Specifically,
$\Lambda^2(-1)$ is equal to $R_{2}(\pi/4)\Bigl(R(-\pi/4)\otimes
R(-\pi/4)\Bigr)$ up to an overall phase. If we add the Hadamard gate $H$, we
obtain all Clifford (symplectic) gates. The Hadamard gate is realized by this
adaptive procedure. We take a qubit in an arbitrary state
$|\psi\rangle=c_{0}|0\rangle+c_{1}|1\rangle$ and supplement it with a
$|+\rangle$ ancilla. Then we apply $\Lambda^2(-1)$ and measure the first qubit
in the dual basis. The second qubit now contains $H|\psi\rangle$ or
$\sigma^{x}H|\psi\rangle$, depending on the measurement outcome. In the second
case, we repeat the procedure $2,4,6,\ldots$ more times until we achieve the
desired result. Using the Clifford gates and the ability to create copies of
$|\xi\rangle=R(\pi/8)|+\rangle$, one can perform quantum
computation. Furthermore, if the Clifford gates are exact, the ancillary state
$|\xi\rangle$ only needs to be prepared with fidelity $F>0.93$~\cite{BK}. This
gives more that $50\%$ tolerance for choosing the parameter $u\approx\pi/8$ in
$R(u)=\exp(iu\sigma^z)$.

\begin{figure}
\centerline{\begin{tabular}{c@{\qquad}c@{\qquad}c}
\includegraphics[scale=0.6]{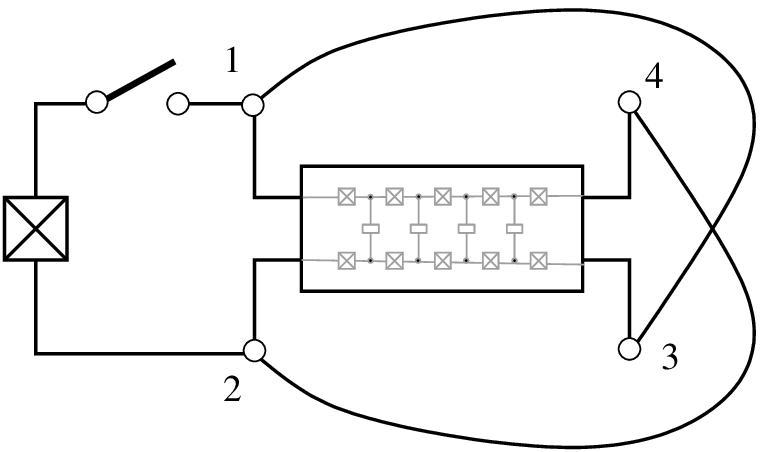} &
\includegraphics[scale=0.6]{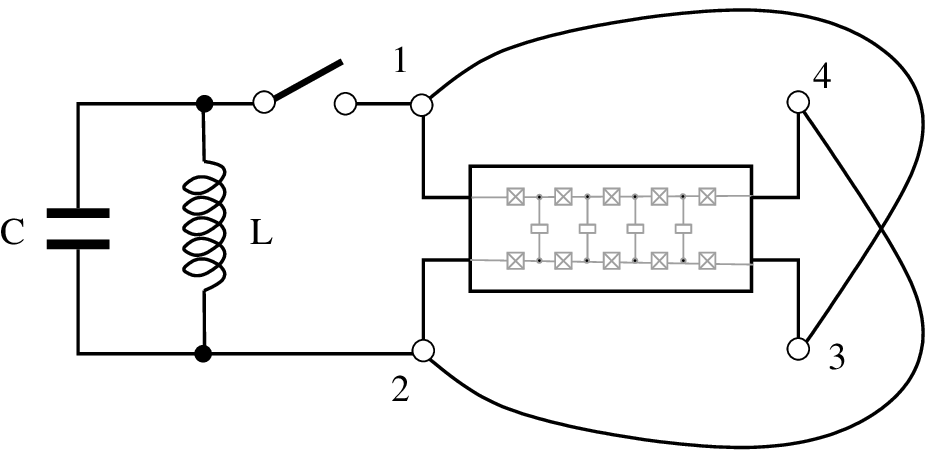} &
\includegraphics[scale=0.6]{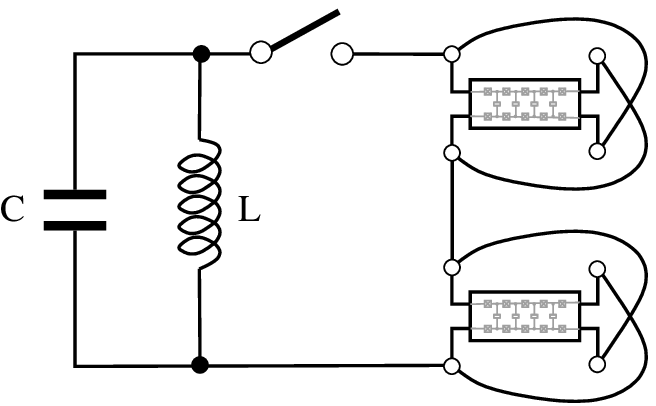} \\
a) & b) & c)
\end{tabular}}
\caption{Realization of the gate $R(u)$ for arbitrary $u$ (a); the
fault-tolerant realization of $R(\pi/4)$ (b) and $R_2(\pi/4)$ (c).}
\label{fig_gate}
\end{figure}

The implementation of measurements is described above. One can readily see
that it is fault-tolerant since the measured observable (i.e., the
superconducting phase or the offset charge) is as unlikely to change during
the process as in an isolated qubit. The gate $R(u)$ is realized by connecting
the leads by a Josephson junction for a certain time period,\footnote{The
required switch may be implemented as a series of SQUIDs controlled by
magnetic field~\cite{HAA}.} Fig.~\ref{fig_gate}a. This procedure is generally
sensitive to random variations of the time interval and the strength of the
Josephson coupling.

A fault-tolerant implementation of $R(\pi/4)=\sqrt{i}\Lambda(-i)$ is based on
a mathematical idea from Ref.~\cite{oscqubit}. The circuit is shown
schematically in Fig.~\ref{fig_gate}b: the $0$-$\pi$ qubit is connected to an
ultraquantum $LC$-oscillator (with $r\bydef(e^2/h)\sqrt{L/C}\gg 1$) for a
certain time period $\tau$. The operation of this gate may be described in
terms of the superconducting phase difference $\theta$ across the inductor. It
is a real variable (not identified modulo $2\pi$) because there are no phase
slips in the inductor. Initially, the oscillator is in its ground state
characterized by a Gaussian wave function $\psi_{0}(\theta)$; note that
$\langle\theta^{2}\rangle\sim r\gg 1$.  Once the circuit is closed, the
quantum evolution is governed by the effective Hamiltonian
\begin{equation}
H_L=(\hbar^2/8e^2)L^{-1}\theta^{2},
\end{equation}
where $\theta$ takes on multiples of $2\pi$ if the qubit is in the state
$|0\rangle$, or on values of the form $2\pi(n+1/2)$ if the qubit is in the
state $|1\rangle$. Thus the wave function $\psi(\theta)$ has the form of a
grid: it consists of narrow peaks at the said locations. If $\tau=8L(e^2/h)$,
then all peaks with $\theta=2\pi n$ pick up no phase and all peaks with
$\theta=2\pi(n+1/2)$ are multiplied by $-i$. Thus the gate $\Lambda(-i)$ is
effectively applied to the qubit state, not entangling it with the
oscillator. Some conditions should be met for this scheme to work. The closing
and breaking of the circuit should occur smoothly enough so that no excitation
is produced in the switch itself, but faster that the $LC$ oscillation
period. The latter (antiadiabatic) condition is needed to prevent the
squeezing of the state to just one or two peaks with the smallest energy. If
these requirements are fulfilled, the qubit is transformed into a Gaussian
grid state, i.e., a superposition of peaks with a broad Gaussian envelope.  It
is important that each peak is narrow and the Fourier image of the envelope is
also narrow; such states are known to have good error-correcting
properties~\cite{oscqubit}. Indeed, a bit flip is unlikely because the peaks
from the $|0\rangle$ and $|1\rangle$ grids do not overlap, and dephasing is
suppressed because $|+\rangle$ and $|-\rangle$ correspond to disjoint grids in
the momentum space. If the protocol is not followed exactly but with a small
error, it will mainly result in oscillations in the $LC$ circuit after the
cycle is complete, leaving the qubit unaffected. The gate $R_{2}(\pi/4)$ is
implemented similarly: one just needs to connect the two $0$-$\pi$ qubits in
series,~Fig.~\ref{fig_gate}c.

Summing up, we have described a concrete design that belongs to a class of
$0$-$\pi$ superconducting qubits. It consists of a current mirror device with
the four leads connected diagonally. This design makes it possible to perform
a measurement with respect to the dual basis, using whatever technique that is
suitable to measure the offset charge of a capacitor. We also propose a
fault-tolerant scheme, including a universal gate set and their schematic
implementations. This scheme can be used with any type of $0$-$\pi$
superconducting qubits.

I thank Lev Ioffe, John Preskill, Michael Freedman, and Gil Rafael for helpful
and encouraging discussions. I acknowledge the support by ARO under Grants
No.\ W911NF-04-1-0236 and W911NF-05-1-0294, and by NSF under Grant No.\
PHY-0456720.


\begin{thebibliography}{99}

\bibitem{NPT} Y.\,Nakamura, Y.\,A.\,Pashkin, J.\,S.\,Tsai, ``Coherent control
of macroscopic quantum states in a single-Cooper-pair box'', \textit{Nature}
\textbf{398}, 786-788 (1999), \texttt{cond-mat/9904003}.

\bibitem{quantronium} D.\,Vion, A.\,Aassime, A.\,Cottet, P.\,Joyez,
H.\,Pothier, C.\,Urbina, D.\,Esteve, M.\,H.\,Devoret, ``Manipulating the
quantum state of an electrical circuit'', \textit{Science} \textbf{296},
886-889 (2002), \texttt{cond-mat/0205343}.

\bibitem{Yu_at_al} Y.\,Yu, S.\,Y.\,Han, X.\,Chu, S.\,I.\,Chu, Z.\,Wang,
``Coherent temporal oscillations of macroscopic quantum states in a Josephson
junction'', \textit{Science} \textbf{296}, 889-892 (2002).

\bibitem{large_qubit} J.\,M.\,Martinis, S.\,Nam, J.\,Aumentado, C.\,Urbina,
 ``Rabi oscillations in a large Josephson-junction qubit'',
 \textit{Phys.\ Rev.\ Lett.}\ \textbf{89}, 117901 (2002).

\bibitem{flux_qubit} I.\,Chiorescu, Y.\,Nakamura, C.\,J.\,P.\,M.\,Harmans,
J.\,E.\,Mooij, ``Coherent quantum dynamics of a superconducting flux qubit'',
\textit{Science} \textbf{299}, 1869-1871 (2003), \texttt{cond-mat/0305461}.

\bibitem{Jgate} T.\,Yamamoto, Yu.\,A.\,Pashkin, O.\,Astafiev, Y.\,Nakamura,
J.\,S.\,Tsai, ``Demonstration of conditional gate operation using
superconducting charge qubits'', \textit{Nature} \textbf{425}, 941-944 (2003),
\texttt{cond-mat/0311067}.

\bibitem{Shor} P.\,W.\,Shor, ``Fault-tolerant quantum computation'',
\textit{37th Annual Symposium on Foundations of Computer Science (FOCS '96)},
p.~56 (1996), \texttt{quant-ph/9605011}.

\bibitem{Kitaev} A.\,Yu.\,Kitaev, ``Fault-tolerant quantum computation by
anyons'', \textit{Annals of Physics} \textbf{303} no.~1, 2--30 (2003),
\texttt{quant-ph/9707021}.

\bibitem{DoucotVidal} B.\,Dou\c{c}ot, J.\,Vidal, ``Pairing of Cooper Pairs in
a Fully Frustrated Josephson Junction Chain'', \textit{Phys.\ Rev.\ Lett.}\
\textbf{88}, 227005 (2002)\texttt{cond-mat/0202115}.

\bibitem{IF02} L.\,B.\,Ioffe, M.\,V.\,Feigel'man, ``Possible realization of an
ideal quantum computer in Josephson junction array'', \textit{Phys.\ Rev.\ B}
\textbf{66}, 224503 (2002),  \texttt{cond-mat/0205186}.

\bibitem{AKN} D.\,V.\,Averin, A.\,N.\,Korotkov, Yu.\,V.\,Nazarov, ``Transport
of electron-hole pairs in arrays of small tunnel junctions'', \textit{Phys.\
Rev.\ Lett.}\ \textbf{66}, 2818--2821 (1991).

\bibitem{SD} H.\,Shimada, P.\,Delsing, ``Current Mirror Effect and Correlated
Cooper-Pair Transport in Coupled Arrays of Small Josephson Junctions'',
\textit{Phys.\ Rev.\ Lett.}\ \textbf{85}, 3253--3256 (2000).

\bibitem{CCL} Mahn-Soo~Choi, M.\,Y.\,Choi, T. Choi, Sung-Ik~Lee, ``Cotunneling
Transport and Quantum Phase Transitions in Coupled Josephson-Junction Chains
with Charge Frustration'', \textit{Phys.\ Rev.\ Lett.}\ \textbf{81}, 4240-4243
(1998), \texttt{cond-mat/9802199}.

\bibitem{LCC} Minchul~Lee, Mahn-Soo~Choi, M.\,Y.\,Choi, ``Quantum phase
transitions and persistent currents in Josephson-junction ladders'',
\textit{Phys.\ Rev.\ B} \textbf{68}, 144506 (2003), \texttt{cond-mat/0302567}.

\bibitem{BK} S.\,Bravyi, A.\,Kitaev, ``Universal Quantum Computation with
ideal Clifford gates and noisy ancillas'', \textit{Phys.\ Rev.\ A}
\textbf{71}, 022316 (2005), \texttt{quant-ph/0403025}.

\bibitem{HAA} D.\,B.\,Haviland, K.\,Andersson, P.\,Agren ``Superconducting and
insulating behavior in one-dimensional Josephson junction arrays'',
\textit{J.\ Low Temp.\ Phys.}\ \textbf{118}, 733-749 (2000),
\texttt{cond-mat/0001143}.

\bibitem{oscqubit} D.\,Gottesman, A.\,Kitaev, J.\,Preskill, ``Encoding a qubit
in an oscillator'', \textit{Phys.\ Rev.\ A} \textbf{64}, 012310 (2001),
\texttt{quant-ph/0008040}.

\end{thebibliography}
\end{document}